\documentclass[correspondence]{IEEEtaes}

\usepackage{amsmath,graphicx}
\usepackage{epsfig} 
\usepackage{dblfloatfix} 
\usepackage{courier}
\usepackage{amssymb}
\usepackage{float}
\usepackage{xcolor}
\usepackage{multirow}
\usepackage{caption}
\usepackage{subcaption}
\usepackage{booktabs}
\newcommand{\ignore}[1]{}

\newcommand{\mc}[1]{\mathcal{#1}}

\newcommand{\bma}[1]{\left[\begin{array}{ #1}}
\newcommand{\ema}{\end{array}\right]}

\DeclareMathAlphabet{\mbf}{OT1}{ptm}{b}{n}
\newcommand{\mbs}[1]{{\boldsymbol{#1}}}

\newcommand{\mbfbar}[1]{{\bar{\mbf{#1}}}}
\newcommand{\mbfhat}[1]{{\hat{\mbf{#1}}}}

\def\fdotb{{\raisebox{-0.6ex}{ \kern0.2ex\raisebox{0.8ex}{\tiny $\hspace*{-1ex}\circ$}}}}
\def\fddotb{{\raisebox{-0.6ex}{ \kern0.2ex\raisebox{0.8ex}{\tiny $\hspace*{-1ex}\circ\circ$}}}}

\newcommand{\f}{\frac}

\newcommand{\dee}{\textrm{d}}

\newcommand{\trans}{{\ensuremath{\mathsf{T}}}} 
\newcommand{\utimes}{ {\raisebox{-0.6ex}{ \kern-1.0ex\raisebox{0.6ex}{ \small $\mathsf{v}$}}} } %
\newcommand{\beq}{\begin{equation}}
\newcommand{\eeq}{\end{equation}}
\newcommand{\bdis}{\begin{displaymath}}
\newcommand{\edis}{\end{displaymath}}
\newcommand{\beqarray}{\begin{eqnarray}}
\newcommand{\eeqarray}{\end{eqnarray}}
\newcommand{\beqarraynn}{\begin{eqnarray*}}
\newcommand{\eeqarraynn}{\end{eqnarray*}}
\newcommand{\balign}{\begin{align}}
\newcommand{\ealign}{\end{align}}
\newcommand{\balignnn}{\begin{align*}}
\newcommand{\ealignnn}{\end{align}}

\renewcommand{\theenumii}{\arabic{enumii}}
\makeatletter
\renewcommand{\p@enumii}{\theenumi.}
\makeatother

\usepackage{arydshln}

\jvol{XX}
\jnum{XX}
\jmonth{XXXXX}
\paper{1234567}
\pubyear{2023}
\doiinfo{TAES.2023.Doi Number}

\newcommand{\Lim}[1]{\raisebox{0.5ex}{\scalebox{0.8}{$\displaystyle \lim_{#1}\;$}}}

\newcommand{\diff}[1]{\textcolor{black}{#1}}
\newcommand{\ms}[1]{\textcolor{black}{#1}}

\begin{document}

\sptitle{Correspondence}
\title{Reducing Two-Way Ranging Variance by Signal-Timing Optimization}

\author{Mohammed Ayman Shalaby}
\member{Student Member, IEEE}
\affil{McGill University, Montreal, QC, Canada} 

\author{Charles Champagne Cossette}
\member{Student Member, IEEE}
\affil{McGill University, Montreal, QC, Canada} 

\author{James Richard Forbes}
\member{Member, IEEE}
\affil{McGill University, Montreal, QC, Canada} 

\author{Jerome Le Ny}
\member{Senior Member, IEEE}
\affil{Polytechnique Montreal, Montreal, QC, Canada} 

\receiveddate{
This work was supported by the NSERC Alliance Grant program, the NSERC Discovery Grant program, the CFI JELF program, and FRQNT Award 2018-PR-253646.}

\corresp{{\itshape (Corresponding author: M. A. Shalaby)}.}

\authoraddress{\textbf{Mohammed A. Shalaby}, \textbf{Charles C. Cossette}, and \textbf{James R. Forbes} are with the Mechanical Engineering Department, McGill University, Montreal, QC  H3A 0C3, Canada (e-mails: mohammed.shalaby@mail.mcgill.ca, charles.cossette@mail.mcgill.ca, james.richard.forbes@mcgill.ca). \textbf{Jerome Le Ny} is with the Electrical Engineering Department, Polytechnique Montreal, Montreal, QC H3T 1J4, Canada (e-mail: jerome.le-ny@polymtl.ca).}

\markboth{CORRESPONDENCE}{}
\maketitle

\begin{abstract}
  Time-of-flight-based \ms{ranging} among transceivers with different clocks requires protocols that accommodate varying rates of the clocks. Double-sided two-way ranging (DS-TWR) is widely adopted as a standard protocol due to its accuracy; however, the precision of DS-TWR has not been clearly addressed. In this paper, an analytical model of the variance of DS-TWR is derived as a function of the user-programmed response delays, \ms{which is then compared to the \emph{Cramer-Rao Lower Bound} (CRLB)}. This is then used to formulate an optimization problem over the response delays in order to maximize the information gained from range measurements. The derived analytical variance model and optimized protocol are validated experimentally with 2 ranging UWB transceivers, where 29 million range measurements are collected.
\end{abstract}


\section{Introduction}
\label{sec:intro}

A common requirement for real-time localization systems (RTLS) is a source of distance or \emph{range} measurements between different bodies, which motivated the adoption of the IEEE 802.15.4a standard \cite{802-15-4a} for radio-frequency systems. Range measurements are obtained by measuring the time-of-flight (ToF) of signals between two transceivers, which requires accurate timestamping of transmission and reception of signals at both transceivers. However, this is not straightforward due to the transceivers' clocks running at different rates, thus introducing a time-varying offset between the clocks \cite{bensky2007, sahinoglu2008, Navratil2022}. The rate of change of this clock offset is hereinafter referred to as the \emph{clock skew}.

Clock offsets and skews between different transceivers introduce biases in the range measurements \cite{Kwak2010, Neirynck2017, shalaby2023}, which are addressed by the ranging protocols presented in the IEEE 802.15.4z standard \cite{802-15-4z}. A commonly used protocol is \emph{two-way ranging} (TWR), which relies on averaging out two ToF measurements in order to negate the effect of the clock offset. This is widely adopted in ultra-wideband (UWB)-based ranging \cite{shalaby2023, Hepp2016a, Laadung2022}, and is also used in underwater applications utilizing acoustic position systems \cite{Vickery1998}, distance measuring equipment (DME) in aviation navigation \cite{Lo2013}, and other radio systems such as Zigbee \cite{Panta2019}.

\begin{figure}[t!]
    \centering
    \begin{subfigure}[t]{0.46\columnwidth}
        \centering
        \includegraphics[width=\columnwidth]{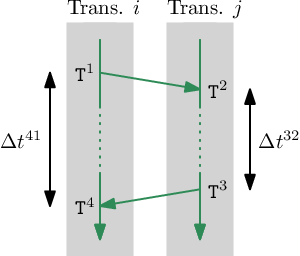}
        \caption{SS-TWR.}
        \label{fig:ss_twr}
    \end{subfigure}%
    ~ \hspace{2pt}
    \begin{subfigure}[t]{0.52\columnwidth}
        \centering
        \includegraphics[width=\columnwidth]{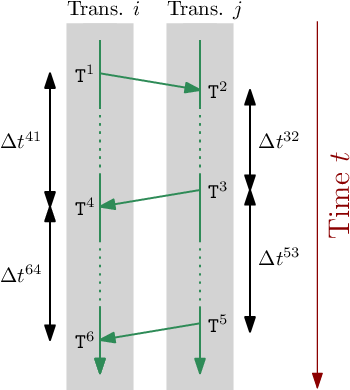}
        \caption{DS-TWR.}
        \label{fig:ds_twr}
    \end{subfigure}
    \caption{Timeline schematics for two transceivers $i$ and $j$ showing the different TWR ranging protocols, where $\mathtt{T}^\ell$ denotes the $\ell^\mathrm{th}$ timestamp for a TWR instance and $\Delta t^{k \ell} \triangleq \mathtt{T}^k - \mathtt{T}^\ell$. \diff{The red arrow indicates the passage of time.}}
    \label{fig:twr}
  \end{figure}

This paper focuses on two variants of the TWR protocol, particularly the single-sided TWR (SS-TWR) and the double-sided TWR (DS-TWR) protocols presented in \cite{shalaby2023}, both shown in Figure~\ref{fig:twr}. Despite requiring an additional message transmission, the main motivation behind DS-TWR as compared to SS-TWR is to correct the clock-skew-dependent bias, which improves the accuracy of the measurements \cite{Neirynck2017, shalaby2023}. \ms{DS-TWR can also be used in the correction of other sources of error, such as the warm-up error \cite{Sidorenko2021}}. Nonetheless, the \emph{precision} of DS-TWR measurements as compared to SS-TWR measurements is a less commonly-addressed topic, where precision is typically measured by the variance of the range measurements. The variance of TWR measurements has been derived analytically as an approximate function of the true range \cite{Jourdan2008, Navratil2019}, \ms{the low-level features of the signal such as the pulse shape \cite{Guvenc2008}, the surrounding environment \cite{Jourdan2008}}, or experimentally for some fixed timing intervals \cite{Sang2019}. 

The main focus of this paper is to extend the comparison between SS-TWR and DS-TWR measurements to include the variance \ms{as a function of the timing delays} in between message transmissions as shown in Figure~\ref{fig:twr}, which allows optimizing signal timing in DS-TWR to improve precision. \ms{Currently, the length of timing delays is arbitrarily chosen; for example, the default DS-TWR code for the commonly-used DW1000 UWB modules \cite{dw1000} appears to include predetermined timing delays without any justification. This paper therefore presents an easily-implementable approach to setting these delays to improve the precision of the range measurements, which in the case of the DW1000 modules is as simple as changing one number in the default code.}

\noindent The contributions of this paper are as follows.
\begin{itemize}
  \item Deriving an analytical model of the variance of SS-TWR and DS-TWR as a function of the timing of message transmissions. 
  \item \ms{Comparing the derived DS-TWR analytical variance to the \emph{Cramer-Rao Lower Bound} (CRLB).}
  \item Formulating an optimization problem for DS-TWR as a function of the signal timings to maximize the information collected in one unit of time. 
  \item Analyzing the effect of relative motion during ranging for DS-TWR.
  \item Validating experimentally the analytical model and the optimization procedure using static UWB transceivers. 
\end{itemize}

The remainder of this paper is organized as follows. \diff{After introducing the notation and assumptions in this paper, the analytical model of the variance and the mean squared error (MSE) of TWR measurements are derived in Section~\ref{sec:var}, and the former compared to the CRLB in Section~\ref{sec:crlb}. The timing-optimization problem is formulated in Section~\ref{sec:opt}}, and experimental validation is then shown in Section~\ref{sec:exp}. 

\subsection{\ms{Notation}}

The $i^\text{th}$ time instance in a TWR transaction is denoted $\mathtt{T}^i \in \mathbb{R}$ as shown in Figure \ref{fig:twr}, and $\mathtt{T}^i_j$ denotes the $i^\text{th}$ time instance as timestamped by Transceiver $j$. The length of time between two time instances $\ell$ and $k$ is denoted $\Delta t^{k\ell} \triangleq \mathtt{T}^k - \mathtt{T}^\ell$. These can also be resolved in a Transceiver's clock, such as $\Delta t^{k\ell}_j = \mathtt{T}^k_j - \mathtt{T}^\ell_j$. The ToF between Transceivers $i$ and $j$ is denoted $t_\mathrm{f}$, and an estimate of the ToF is denoted $\hat{t}_\mathrm{f}$. The time-varying \emph{clock offset} $\tau_i(t)$ of Transceiver $i$ is defined as \diff{$\tau_i(t)~\triangleq t_i(t)~-~t$}, where $t_i(t)$ is the time $t$ resolved in Transceiver $i$'s clock. The \emph{clock skew} of Transceiver $i$ is denoted $\gamma_i$ and is defined as $$\diff{\gamma_i(t) \triangleq \dot{\tau}_i(t) = \Lim{\Delta t \to 0} \f{1}{\Delta t} \left(\tau_i(t + \Delta t) - \tau_i(t)\right).}$$

\subsection{Assumptions} \label{subsec:assumptions}

\diff{
    It is assumed that clock skews are constant during a ranging transaction, which is a common assumption in localization applications due to the clocks' slow dynamics \cite{sahinoglu2008, Neirynck2017}. Therefore, under a first-order approximation, the clock offset at two different time instants separated by $\Delta t$ can be approximately related by}
    \beq
        \diff{\tau_i(t + \Delta t) \approx \tau_i(t) + \gamma_i(t)\Delta t. \label{eq:clock_skew}}
    \eeq 

\diff{The IEEE 802.15.4a standard for radio-frequency systems accommodates for clock skews up to $\pm20$ parts-per-million (ppm) \cite{802-15-4a,Neirynck2017}, which is the order of the worst-case clock skew assumed in this paper. Therefore, throughout this paper, it is assumed that $\gamma_i \ll 1$.} 

\diff{Furthermore, it is assumed that $t_\mathrm{f} \ll \Delta t^{32}$ and $t_\mathrm{f} \ll \Delta t^{53}$, which is a reasonable assumption for short-range systems up to the order of tens or hundreds of meters. For example, 30 meters is equivalent to a ToF of 100 ns, while $\Delta t^{32}$ and $\Delta t^{53}$ are typically in the order of milliseconds or hundreds of microseconds to allow sufficient processing time in between transmitted messages. This assumption is less accurate for long-distance ranging, for example in the order of kilometres or more, which nonetheless is not common in UWB ranging.}

\diff{Due to the aforementioned assumptions, approximations such as $\gamma_i \Delta t^{41}\approx~\gamma_i\Delta t^{32}$ are made throughout this paper. This follows from the term $\gamma_i t_f$ being much smaller than $\gamma_i \Delta t^{32}$, since $t_\mathrm{f} \ll \Delta t^{32}$ and $\gamma_i t_f$ corresponds to a value that is in the order of tens or hundreds of micrometers when multiplied by the speed of light, and can thus be neglected.} 
    


\section{TWR Variance} \label{sec:var}

\color{black}

\subsection{\ms{Modelling the Timestamps}}

The measurement models for the timestamps recorded by Transceivers $i$ and $j$ in Figure \ref{fig:twr} are first presented under the assumption that all transceivers are static. The noisy timestamps recorded by Transceiver $i$ in Figure \ref{fig:ss_twr} are modelled as
\begin{align}
    \mathtt{T}^1_i &= \mathtt{T}^1 + \tau_i(\mathtt{T}^1) + \eta^1, \label{eq:t1}\\
    \mathtt{T}^4_i &= \mathtt{T}^1 + 2t_\mathrm{f} + \Delta t^{32} + \tau_i(\mathtt{T}^4) + \eta^4,
\end{align}
where $\eta^\ell$ is random noise on the $\ell^\text{th}$ measurement. All random noise variables on timestamps are assumed to be mutually independent, zero-mean, and with the same variance $\sigma^2$. 

Similarly, the noisy timestamps recorded by Transceiver $j$ in Figure \ref{fig:ss_twr} are modelled as
\begin{align}
  \mathtt{T}^2_j &= \mathtt{T}^1 + t_\mathrm{f} + \tau_j(\mathtt{T}^2) + \eta^2, \\
  \mathtt{T}^3_j &= \mathtt{T}^1 + t_\mathrm{f} + \Delta t^{32} + \tau_j(\mathtt{T}^3) + \eta^3, \label{eq:t3}
\end{align}
while the additional timestamps when performing DS-TWR as in Figure \ref{fig:ds_twr} are modelled as
\begin{align}
  \mathtt{T}^5_j &= \mathtt{T}^1 + t_\mathrm{f} + \Delta t^{32} + \Delta t^{53} + \tau_j(\mathtt{T}^5) + \eta^5, \\
  \mathtt{T}^6_i &= \mathtt{T}^1 + 2t_\mathrm{f} + \Delta t^{32} + \Delta t^{53} + \tau_i(\mathtt{T}^6) + \eta^6. \label{eq:t6}
\end{align}

\diff{
Based on the aforementioned assumptions in Section \ref{sec:intro} and the relation in \eqref{eq:clock_skew}, the offsets in \eqref{eq:t1}-\eqref{eq:t6} can be written as a function of the clock skew, the clock offsets at $\mathtt{T}^1$, and the time delays $\Delta t^{32}$, $\Delta t^{53}$. For example, 
\begin{align*}
    \tau_i(\mathtt{T}^6) &= \tau_i(\mathtt{T}^1 + \Delta t^{61}) \\
    &\approx \tau_i(\mathtt{T}^1) + \gamma_i(\mathtt{T}^1) \Delta t^{61} \\
    &\approx \tau_i(\mathtt{T}^1) + \gamma_i(\mathtt{T}^1) (\Delta t^{32} + \Delta t^{53}).
\end{align*}
A similar process can be followed for the other offsets. The remainder of this paper will oftentimes drop the explicit dependence on $\mathtt{T}^1$ from the notation for brevity.
}

\subsection{\ms{Deriving SS-TWR Variance}}

A SS-TWR ToF estimate $\hat{t}_\mathrm{f}$ of the true ToF $t_\mathrm{f}$ can be computed from \eqref{eq:t1}-\eqref{eq:t3} as
\begin{align*}
    \hat{t}_f^{\text{ss}} &= \f{1}{2} (\Delta t_i^{41} - \Delta t_j^{32}) \\
    &= \f{1}{2} \left( 2t_\mathrm{f} + \Delta t^{32} + \gamma_i \Delta t^{41} + \eta^{41} - (1+\gamma_j)\Delta t^{32} - \eta^{32} \right) \\
    &\approx t_\mathrm{f} + \f{1}{2} \gamma_{ij} \Delta t^{32} + \f{1}{2} (\eta^{41} - \eta^{32}),
\end{align*}
where $\gamma_{ij} \triangleq \gamma_{i} - \gamma_{j}$, $\eta^{k \ell} \triangleq \eta^{k} - \eta^{\ell}$, and $\gamma_i \Delta t^{41} \approx \gamma_i \Delta t^{32}$. Defining the SS-TWR ToF error as $e^{\text{ss}} \triangleq \hat{t}_f^{\text{ss}} - t_\mathrm{f}$, the expected value of the error is 
\begin{align}
    \mathbb{E} [e^\text{ss}] = \f{1}{2} \gamma_{ij} \Delta t^{32}, \label{eq:ss_mean}
\end{align}
which means that $\hat{t}_f^{\text{ss}}$ is in fact a biased measurement of $t_\mathrm{f}$. Meanwhile, the covariance on the measurement is 
\begin{align}
  \mathbb{E} [\left(e^\text{ss} - \mathbb{E} [e^\text{ss}]\right)^2] = \sigma^2. \label{eq:ss_var}
\end{align}

\subsection{\ms{DS-TWR Variance}}

The main motive behind using DS-TWR protocols rather than SS-TWR protocols is to correct the clock-skew-dependent bias in \eqref{eq:ss_mean}. As shown in \cite[Eq. (6)]{shalaby2023}, the DS-TWR ToF estimate from \eqref{eq:t1}-\eqref{eq:t6} can be modelled as 
\begin{align}
    \hat{t}_f^{\text{ds}} &= \f{1}{2} \left( \Delta t_i^{41} - \f{\Delta t_i^{64}}{\Delta t_j^{53}}\Delta t_j^{32} \right) \label{eq:dstwr} \\
    &\approx t_\mathrm{f} + \f{1}{2} \left( \f{\Delta t^{32}}{\Delta t^{53}} (\eta^{53} - \eta^{64}) + \eta^{41} - \eta^{32} \right), \label{eq:dstwr_model}
\end{align}
where the approximations $\gamma_i t_\mathrm{f} \approx 0$ and $\gamma_i \eta \approx 0$ are used since the clock skew, time-of-flight, and timestamping noise are all small. Defining the DS-TWR ToF error as $e^{\text{ds}} \triangleq \hat{t}_f^{\text{ds}} - t_\mathrm{f}$, the expected value of the error is 
\begin{align*}
    \mathbb{E} [e^\text{ds}] = 0,
\end{align*}
meaning that unlike $\hat{t}_f^{\text{ss}}$, the estimate $\hat{t}_f^{\text{ds}}$ is unbiased. 

Having addressed the accuracy of the measurements for SS-TWR and DS-TWR, it might appear that DS-TWR should always be used. However, the choice of ranging protocol should also depend on the precision of the measurements. By manipulating \eqref{eq:dstwr_model}, the covariance of $\hat{t}_f^{\text{ds}}$ can be found to be of the form
\begin{align}
  \mathbb{E} [\left(e^\text{ds} - \mathbb{E} [e^\text{ds}]\right)^2] = \sigma^2 \left( 1 + \f{\Delta t^{32}}{\Delta t^{53}} + \left(\f{\Delta t^{32}}{\Delta t^{53}}\right)^2 \right). \label{eq:var_ds}
\end{align}
Therefore, 
the variance of DS-TWR measurements is greater than SS-TWR measurements, and approaches the variance of SS-TWR as $\Delta t^{32} \to 0$ and/or $\Delta t^{53} \to \infty$. The $\Delta t^{32} \to 0$ condition is due to the effect of the length of $\Delta t^{32}$ on the bias, and the $\Delta t^{53} \to \infty$ condition is due to the fact that the ratio $\f{\Delta t^{64}}{\Delta t^{53}}$ is being used to obtain a clock-skew measurement, and the longer the $\Delta t^{53}$ interval is the greater the signal-to-timestamping-noise ratio.

\subsection{\diff{Mean Squared Error of SS-TWR and DS-TWR}}

\diff{Knowing the mean-bias and the variance of the ToF estimates for SS-TWR and DS-TWR allows computing the mean squared error (MSE) of the estimates. The MSE of SS-TWR from \eqref{eq:ss_mean} and \eqref{eq:ss_var} is
\begin{align}
    \mathbb{E}\left[ (e^\text{ss})^2 \right] &= \mathbb{E}\left[ (e^\text{ss} - \mathbb{E} [e^\text{ss}])^2 \right] + \mathbb{E} [e^\text{ss}]^2 \nonumber \\
    &= \sigma^2 + \f{1}{4} \gamma_{ij}^2 (\Delta t^{32})^2, \label{eq:ss_mse}
\end{align} 
and the MSE of DS-TWR is the same as \eqref{eq:var_ds} since the estimate is unbiased.
}

\begin{figure}[t]
    \centering
    \includegraphics[width=\columnwidth]{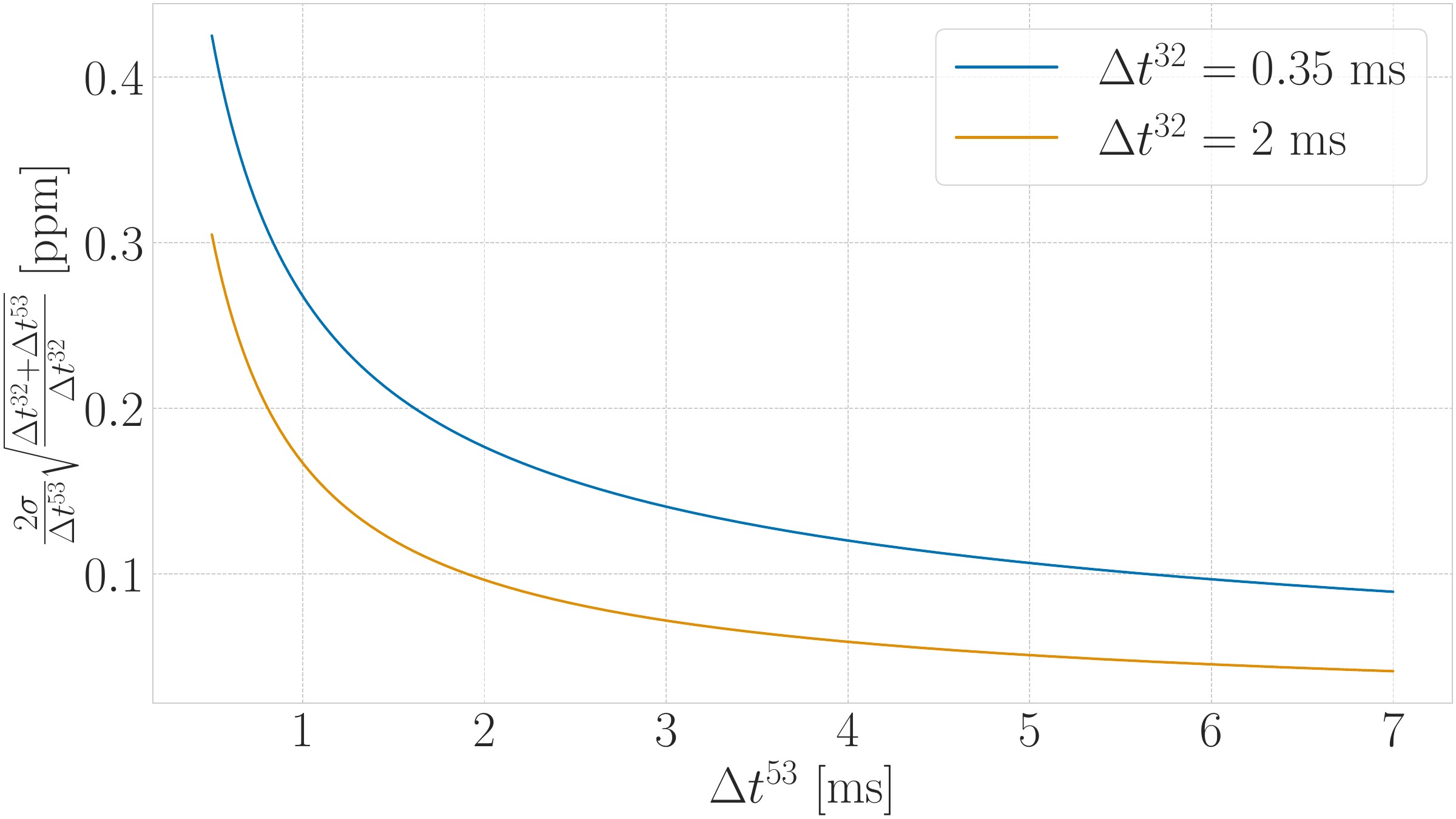}
    \caption{\diff{The value of the right-hand-side of \eqref{eq:gamma_condition} for different values of $\Delta t^{32}$ and $\Delta t^{53}$ when $\sigma = 0.0682$ ns. These curves represent a lower bound for the magnitude of the clock skew $\gamma_{ij}$ for which the MSE of DS-TWR is lower than the MSE of SS-TWR.}}
    \label{fig:mse_comparison}
\end{figure}

\diff{
Therefore, the MSE of DS-TWR is lower than the MSE of SS-TWR when 
\beq
    \sigma^2 \left( 1 + \f{\Delta t^{32}}{\Delta t^{53}} + \left(\f{\Delta t^{32}}{\Delta t^{53}}\right)^2 \right) < \sigma^2 + \f{1}{4} \gamma_{ij}^2 (\Delta t^{32})^2, \nonumber
\eeq
which can also be written as
\beq
    \vert \gamma_{ij} \vert > \f{2\sigma}{\Delta t^{53}} \sqrt{\f{\Delta t^{32} + \Delta t^{53}}{\Delta t^{32}}}. \label{eq:gamma_condition}
\eeq
The right-hand-side of \eqref{eq:gamma_condition} is plotted in Figure \ref{fig:mse_comparison} for $\sigma = 0.0682$ ns, where the value for $\sigma$ is determined experimentally in Section \ref{sec:exp}. Given that $\vert \gamma_{ij} \vert$ is expected to range between 0 and 40 ppm, it is very likely that the MSE of DS-TWR will be lower than that of SS-TWR, except for highly-accurate clocks with lower skew resulting in lower measurement bias.
}

\section{Cramer-Rao Lower Bound of DS-TWR} \label{sec:crlb}

Given that an analytical model is available for the variance of the range estimate provided by the DS-TWR protocol \eqref{eq:dstwr}, the variance of the DS-TWR estimator can be compared to the CRLB \cite[Chapter 3]{Kay1993}.

When two transceivers are ranging with one another, time instances in global time and offsets of individual clocks remain unknown, and only the time instances in a transceiver's clocks and relative offset between the two Transceivers can be estimated. Therefore, the unknown quantities to be estimated from \eqref{eq:t1}-\eqref{eq:t6} can be summarized in a state vector
\begin{align*}
    \mbf{x} = \bma{cccccc} t_\mathrm{f} & \mc{T} & \tau_{ij} & \gamma_{ij} & \Delta t_j^{32} & \Delta t_j^{53}  \ema^\trans,
\end{align*}
where $\mc{T} = \mathtt{T}^1 + \tau_i(\mathtt{T}^1)$ and $\tau_{ij} \triangleq \tau_i(\mathtt{T}^1) - \tau_j(\mathtt{T}^1)$. Additionally, under the assumption that $\f{\gamma_{ij}}{1 + \gamma_j} \approx \gamma_{ij}$ since $\gamma_j~\ll~1$, it can be shown that
\begin{align*}
    \Delta t_i^{32} \approx (1+\gamma_{ij}) \Delta t_j^{32}, \\
    \Delta t_i^{53} \approx (1+\gamma_{ij}) \Delta t_j^{53}.
\end{align*}
Therefore, the timestamp measurements \eqref{eq:t1}-\eqref{eq:t6} can be written as
\begin{align*}
    \mathtt{T}_i^1(\mbf{x}) &= \mc{T} + \eta^1, \\
    \mathtt{T}_j^2(\mbf{x}) &= \mc{T} + t_\mathrm{f} - \tau_{ij} + \eta^2, \\
    \mathtt{T}_j^3(\mbf{x}) &\approx \mc{T} + t_\mathrm{f} - \tau_{ij} + \Delta t_j^{32} + \eta^3, \\
    \mathtt{T}_i^4(\mbf{x}) &\approx \mc{T} + 2t_\mathrm{f} + (1+\gamma_{ij}) \Delta t_j^{32} + \eta^4, \\
    \mathtt{T}_j^5(\mbf{x}) &\approx \mc{T} + t_\mathrm{f} - \tau_{ij} + \Delta t_j^{32} + \Delta t_j^{53} + \eta^5, \\
    \mathtt{T}_i^6(\mbf{x}) &\approx \mc{T} + 2t_\mathrm{f} + (1+\gamma_{ij}) (\Delta t_j^{32} + \Delta t_j^{53}) + \eta^6,
\end{align*}
where the approximation $\gamma_i t_\mathrm{f} \approx 0$ has been used. The measurement vector can then be written as
\begin{align*}
  \mbf{y}(\mbf{x}) = \bma{cccccc} \mathtt{T}_i^1 & \mathtt{T}_j^2 & \mathtt{T}_j^3 & \mathtt{T}_i^4 & \mathtt{T}_j^5 & \mathtt{T}_i^6 \ema^\trans,
\end{align*}
which is a nonlinear function of the states $\mbf{x}$. Therefore, the measurement Jacobian can be computed as 
\begin{align*}
    \mbf{C} &= \f{\partial \mbf{y}(\mbf{x})}{\partial \mbf{x}} \bigg \vert_\mbfbar{x}\\
    &= \bma{cccccc}
        0 & 1 & 0 & 0 & 0 & 0 \\
        1 & 1 & -1 & 0 & 0 & 0 \\ 
        1 & 1 & -1 & 0 & 1 & 0 \\
        2 & 1 & 0 & \Delta \bar{t}_j^{32} & 1 + \bar{\gamma}_{ij} & 0 \\
        1 & 1 & -1 & 0 & 1 & 1 \\
        2 & 1 & 0 & \Delta \bar{t}_j^{32} + \Delta \bar{t}_j^{53} & 1 + \bar{\gamma}_{ij} & 1 + \bar{\gamma}_{ij}
    \ema,
\end{align*}
where overbars denote the linearization point. Additionally, define a measurement vector covariance $\mbs{\Sigma} \triangleq \sigma^2 \mbf{1}_6$, where $\mbf{1}_6$ is the $6 \times 6$ identity matrix. 

The CRLB states that the covariance of any unbiased estimate $\mbfhat{x}$ of $\mbf{x}$, given the measurements $\mbf{y}(\mbf{x})$ and an additive-Gaussian assumption on the measurement noise, is bounded by \cite[Appendix 3C]{Kay1993}
\begin{align*}
    \mathbb{E} \left[ (\mbf{x} - \mbfhat{x})(\mbf{x} - \mbfhat{x})^\trans \right] \geq (\mbf{C}^\trans \mbs{\Sigma}^{-1} \mbf{C})^{-1}.
\end{align*} 
The minimum variance of the ToF estimate for the given timestamps can then be found by extracting the first component of $(\mbf{C}^\trans \mbs{\Sigma}^{-1} \mbf{C})^{-1}$, which can be found to be
\begin{align*}
    \sigma^2 \f{(\bar{\gamma}_{ij}^2 + 2\bar{\gamma}_{ij} + 2)((\Delta \bar{t}_j^{32})^2 + \Delta \bar{t}_j^{32}\Delta \bar{t}_j^{53} + (\Delta \bar{t}_j^{53})^2)}{2 (\Delta \bar{t}_j^{53})^2}.
\end{align*}
Given that $\bar{\gamma}_{ij}^2 + 2\bar{\gamma}_{ij} \ll 2$, this can be simplified to give exactly \eqref{eq:var_ds}, thus showing that under the aforementioned approximations the DS-TWR estimator is indeed a minimum-variance unbiased estimator.

\color{black}

\section{DS-TWR Timing Optimization} \label{sec:opt}

The \emph{timing delays} $\Delta t^{32}$ and $\Delta t^{53}$ affect the variance of the \ms{range} measurements, the rate of the measurements, and the ranging error due to relative motion between the transceivers. In Section \ref{subsec:optimal_delay}, the choice of delays is motivated as a function of the variance and rate of the measurements while assuming no relative motion between the transceivers. This assumption is then validated in Section \ref{subsec:motion} for the DS-TWR protocol, showing that motion can indeed be neglected when choosing the timing delays.

\subsection{Finding Optimal Timing Delays} \label{subsec:optimal_delay}

Given \eqref{eq:var_ds}, minimizing $\Delta t^{32}$ within the limitations of the system is an obvious choice to reduce the measurement variance. However, it is less clear what the right choice for $\Delta t^{53}$ is, as increasing this second-response delay reduces measurement variance but also reduces the rate of measurements. The choice of $\Delta t^{53}$ is thus application-specific. Most commonly in estimation applications, the goal is to minimize the variance of the estimates, which is achieved by maximizing the \emph{information} obtained from measurements. Therefore, this section poses an information-maximizing (variance-minimizing) optimization problem. 

The amount of information obtained in one unit of time is a function of the variance of the individual measurement and the number of measurements in that unit of time. As a result, the \emph{optimal} delay is one that is long enough to reduce the variance of the individual measurement but short enough to ensure measurements are recorded at a sufficient rate.

The rate of the measurements is dependent on $\Delta t^{32} + \Delta t^{53}$ as well as any further processing \ms{required} to retrieve the range measurements, such as reading the raw timestamps from the registers and computing the range measurement from the raw timestamps. The time taken for computational processing is defined as $\rho$, which is assumed to be constant for the same experimental set-up. Therefore, the time-length of one measurement is $\rho + \Delta t^{32} + \Delta t^{53}$ seconds long. The delay $\Delta t^{32}$ is to be minimized as much as the hardware allows, and $\Delta t^{53}$ is to be optimized as follows. In one second, a total of $\big[\f{1}{\rho + \Delta t^{32} + \Delta t^{53}}\big]$ measurements occur, meaning that, assuming independence, the variance of averaging out the ToF estimates is given as
\begin{align}
    R_\mathrm{avg}(\Delta t^{53}) &\triangleq [\rho + \Delta t^{32} + \Delta t^{53}] R_\mathrm{meas} (\Delta t^{53}), \label{eq:Rt}
\end{align}
where $R_\text{meas} (\Delta t^{53})$ is the variance of the individual measurement given by \eqref{eq:var_ds} for some constant $\Delta t^{32}$. $R_\mathrm{avg}$ is referred to hereinafter as the \emph{averaged uncertainty}, and can be thought of as the inverse of accumulated \emph{information} in one second. The optimal delay $\Delta t^{53^\ast}$ is then found by solving 
\beq
  \Delta t^{53^\ast} = \arg \min_{\Delta t^{53} \in \mathbb{R}} R_\mathrm{avg} (\Delta t^{53}). \label{eq:opt}
\eeq
The derivative of \eqref{eq:Rt} with respect to $\Delta t^{53}$ is
\begin{align}
    \f{\dee R_\mathrm{avg}}{\dee \Delta t^{53}} &= \sigma^2 - (\rho + \Delta t^{32}) \f{\Delta t^{32}}{(\Delta t^{53})^{2}} \sigma^2 \nonumber \\ &\hspace{12pt} - 2(\rho + \Delta t^{32}) \f{(\Delta t^{32})^2}{(\Delta t^{53})^{3}} \sigma^2 - \f{(\Delta t^{32})^2}{(\Delta t^{53})^{2}} \sigma^2, \nonumber
\end{align}
and equating to 0 yields the cubic polynomial
\begin{align}
    0 &= (\Delta t^{53})^3 - \Delta t^{32} (\rho + 2\Delta t^{32}) \Delta t^{53} \nonumber \\ &\hspace{12pt} - 2(\Delta t^{32})^2 (\rho + \Delta t^{32}). \label{eq:optimal_delay}
\end{align}
\ms{This is a ``depressed cubic equation'' that can be solved analytically using Cardano's method, but the analytical solution is omitted here for conciseness. Additionally, this can be solved numerically using standard libraries (such as Bullet's $\operatorname{cubic\_roots}$ function in C++ or NumPy's $\operatorname{roots}$ function in Python).}

\begin{figure}[t]
  \centering
  \begin{subfigure}[t]{\columnwidth}
      \centering
      \includegraphics[width=\columnwidth]{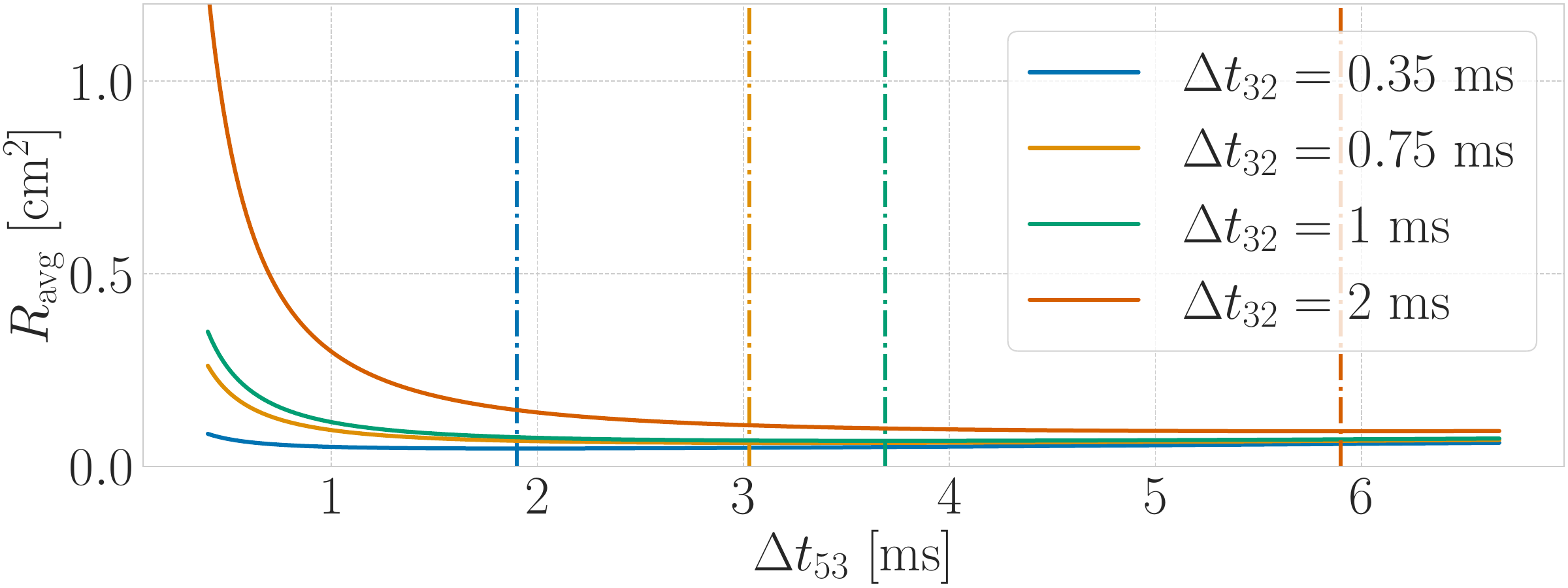}
      \label{fig:rt_vs_t53_theoretical}
  \end{subfigure}%
  \\ 
  \begin{subfigure}[t]{\columnwidth}
      \centering
      \includegraphics[width=\columnwidth]{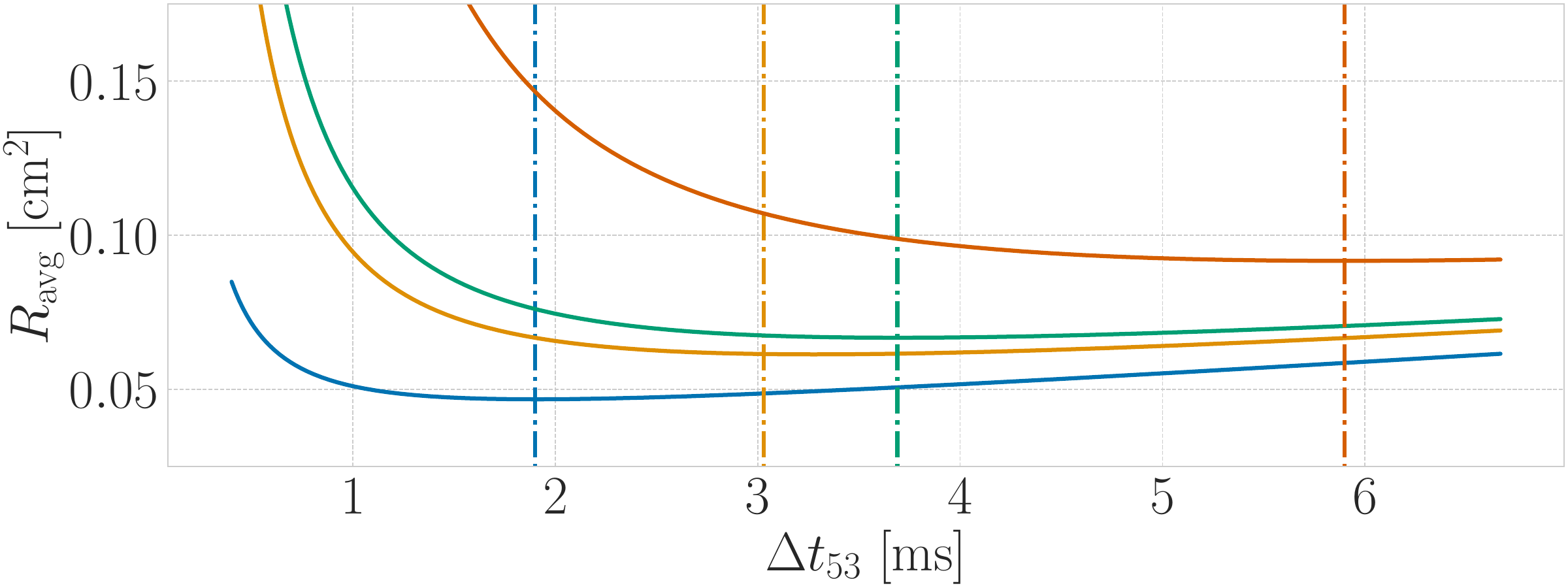}
      \label{fig:rt_vs_t53_theoretical_closeup}
  \end{subfigure}
  \caption{The theoretical averaged variance $R_\mathrm{avg}$ as a function of the delay $\Delta t^{53}$ for 4 different values of $\Delta t^{32}$. All curves use $\rho = 7.2$ ms, which is experimentally determined for the set-up used in Section~\ref{sec:exp}. The vertical dotted lines correspond to the analytically-evaluated minimum of the colour-matched plotted curves. The bottom plot is a close-up view of the top plot. \ms{Note that $R_\text{avg}$ is converted from units of [s]$^2$ to [cm]$^2$ by multiplying with $c^2$ [cm$^2$ / s$^2$], where $c$ is the speed of light, in order to visualize the variance on the range measurements directly.}}
  \label{fig:rt_vs_t53}
\end{figure}

The value for $\rho$ and the minimum value of $\Delta t^{32}$ can be determined experimentally and are both processor and application dependent. As an example where $\rho=7.2$ ms and $\Delta t^{32}=0.35$ ms, the optimal delay can be found analytically to be approximately 1.9 ms using \eqref{eq:optimal_delay}. The averaged variance $R_\mathrm{avg}$ as a function of $\Delta t^{53}$ for $\rho=7.2$ ms at different values of $\Delta t^{32}$ is shown in Figure~\ref{fig:rt_vs_t53}. As expected from \eqref{eq:var_ds}, the averaged variance $R_\text{avg}$ diverges as $\Delta t^{53}$ approaches 0 ms.

\subsection{Relative Motion During Ranging} \label{subsec:motion}

A constant distance throughout ranging is commonly assumed, but this assumption introduces larger errors for longer response delays. To address this, assume the less-restrictive case of no relative acceleration between the transceivers. In this case, the three ToF measurements shown in Figure~\ref{fig:ds_twr} are of different distances, and are related by
\begin{align}
    t_\mathrm{f}^2 &= t_\mathrm{f}^1 + \bar{v} \Delta t^{32}, \qquad
    t_\mathrm{f}^3 = t_\mathrm{f}^2 + \bar{v} \Delta t^{53}, \nonumber
\end{align}
where $t_\mathrm{f}^i$ is the ToF of the $i^\text{th}$ message, $\bar{v} = v / c$, $v$ is the rate of change of the distance between transceivers, and $c$ is the speed of light. Note that motion during the intervals $\Delta t^{32}$ and $\Delta t^{53}$ is addressed since the intervals are in the order of milliseconds. Meanwhile, ToF is much shorter for short range measurements as discussed in Section \ref{sec:intro}, so motion in between time of transmission and reception is negligible.

The computed ToF measurement using the DS-TWR protocol in the absence of clock offsets, skews, and timestamping noise is then
\begin{align*}
    \hat{t}_\mathrm{f}^\mathrm{ds} &= \f{1}{2} \Big( \Delta t^{41} - \f{\Delta t^{64}}{\Delta t^{53}} \Delta t^{32} \Big)\\
    &\ms{= \f{1}{2} \Big( t_\mathrm{f}^1 + \Delta t^{32} + t_\mathrm{f}^2 - \f{\Delta t^{53} + t_\mathrm{f}^3 - t_\mathrm{f}^2}{\Delta t^{53}} \Delta t^{32} \Big)} \\
    &\ms{= \f{1}{2} \Big( 2t_\mathrm{f}^1 + (1+\bar{v})\Delta t^{32} - \f{(1+\bar{v})\Delta t^{53}}{\Delta t^{53}} \Delta t^{32} \Big)} \\
    &= t_\mathrm{f}^1,
\end{align*}
meaning that the computed ToF corresponds to the distance between the transceivers at the beginning of ranging, and the error due to motion is independent from the delays $\Delta t^{32}$ and $\Delta t^{53}$. Therefore, a particular feature of the DS-TWR protocol presented in \cite{shalaby2023} is that the timing optimization can be done without addressing errors due to motion. 

\section{Experimental Evaluation} \label{sec:exp}

\begin{figure}[t!]
  \centering
  \begin{subfigure}[t]{0.4\columnwidth}
      \centering
      \includegraphics[trim={11cm 0cm 11cm 0cm},width=\columnwidth]{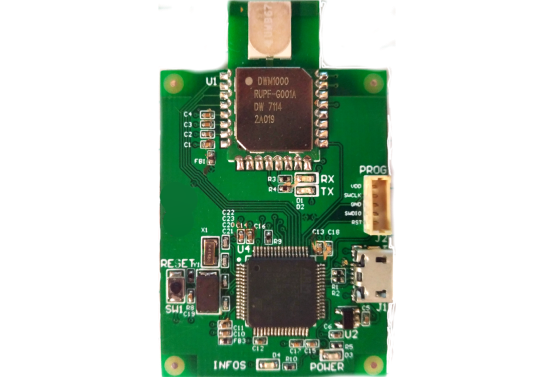}
  \end{subfigure}%
  ~ \hspace{10pt}
  \begin{subfigure}[t]{0.5\columnwidth}
      \centering
      \includegraphics[width=\columnwidth]{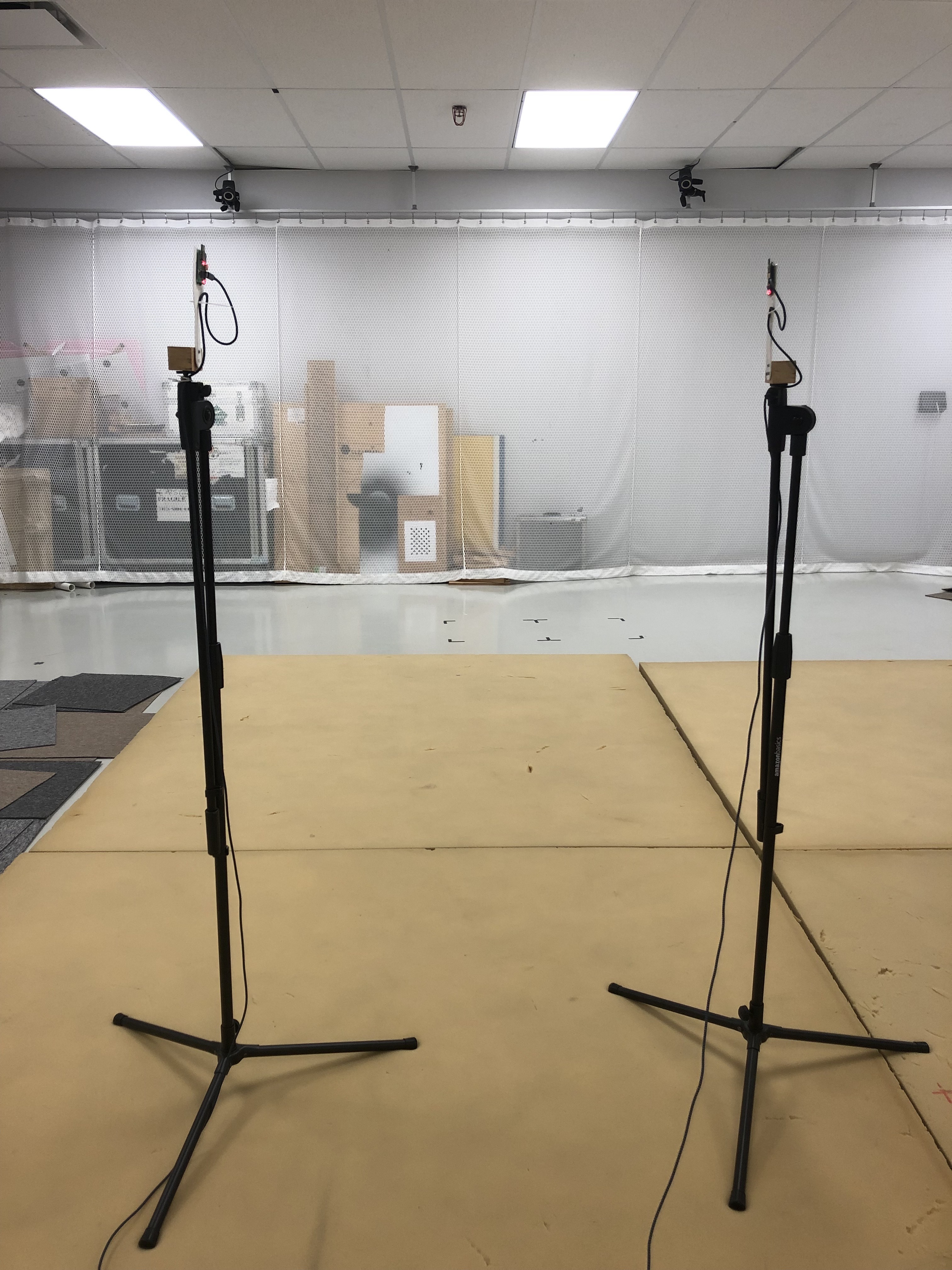}
  \end{subfigure}
  \caption{The experimental set-up. (Left) Custom-built circuit board, using the DWM1000 UWB transceiver. (Right) Two static tripods placed 1.5 metres apart, each holding a UWB transceiver.}
  \label{fig:optimal_delay_experiment}
\end{figure}

\begin{figure}[t]
  \centering
  \begin{subfigure}[t]{\columnwidth}
      \centering
      \includegraphics[width=\columnwidth]{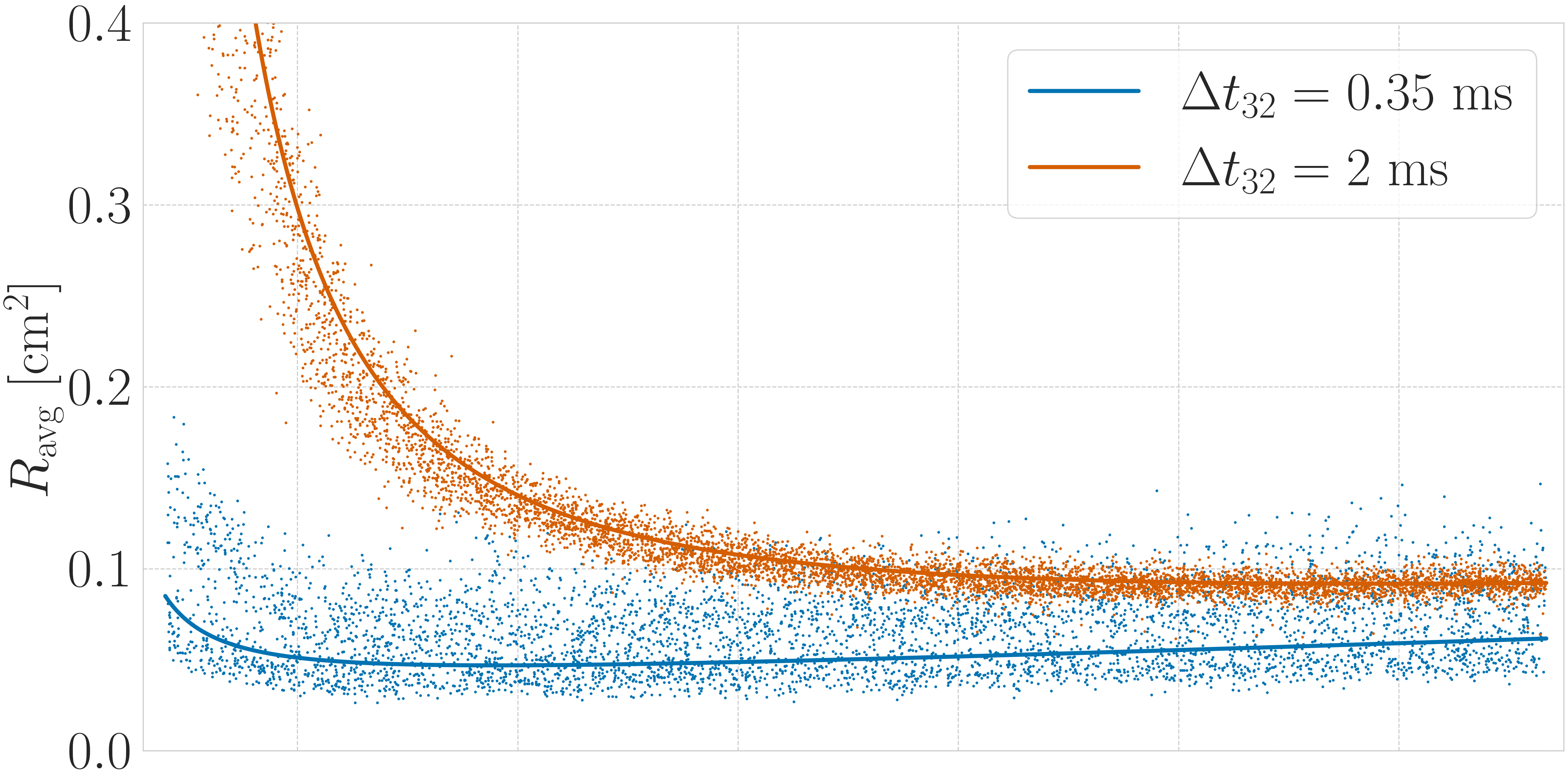}
  \end{subfigure}
  \\ \vspace{12pt}
  \begin{subfigure}[t]{\columnwidth}
      \centering
      \includegraphics[width=\columnwidth]{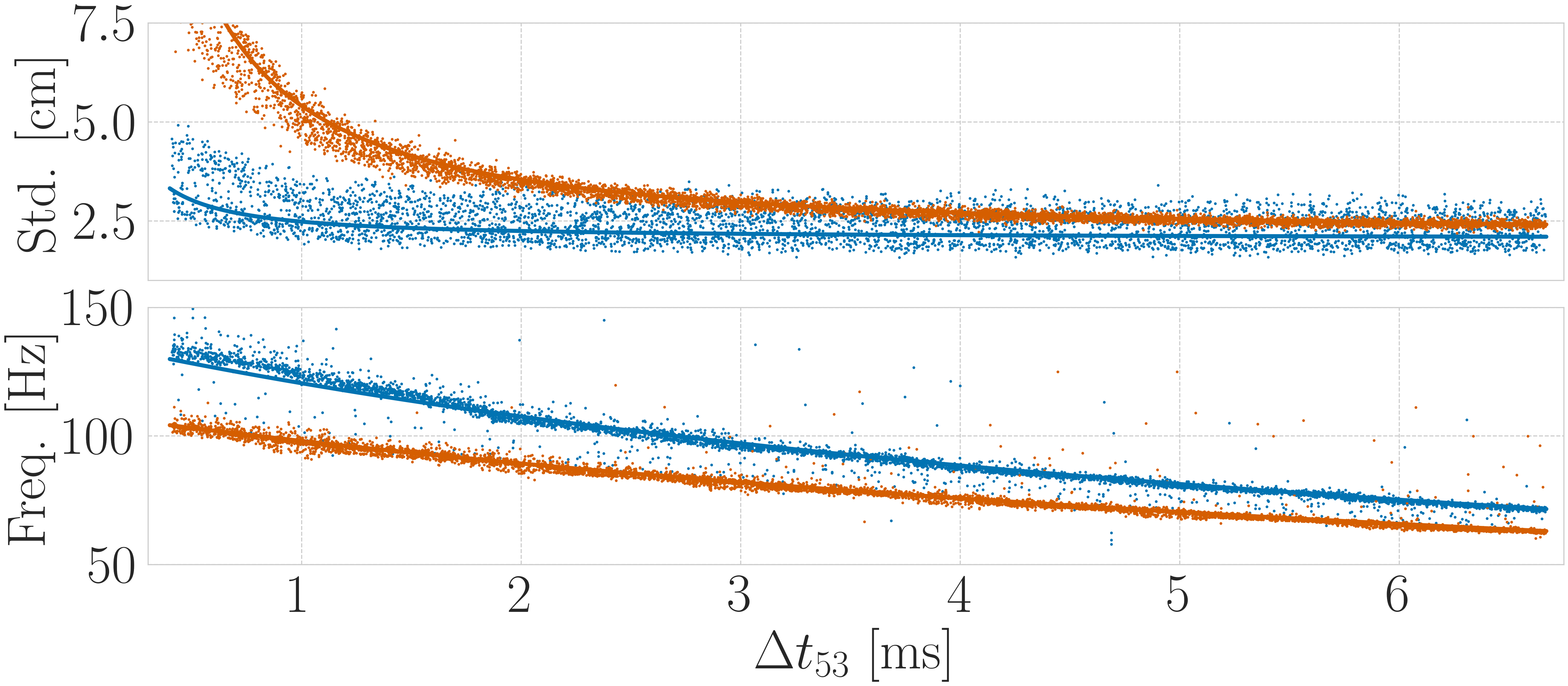}
  \end{subfigure}%
  
  \caption{The theoretical and experimental metrics as they vary with $\Delta t^{53}$. Each point corresponds to one trial of 2500 measurements, and the solid line is the theoretical curve based on the derived analytical models \diff{and the experimentally-computed variance $\sigma^2$}. The solid line matches the experimental readings. The plots show the variation of the averaged variance $R_\mathrm{avg}$ as given by \eqref{eq:Rt}, the standard deviation as given by the square root of \eqref{eq:var_ds}, and the rate of the measurements as a function of $\Delta t^{53}$ for two different values of $\Delta t^{32}$.}
  \label{fig:optimal_delay_metrics}
\end{figure}

To evaluate the effect of the second-response delay $\Delta t^{53}$ on a real system, the following experiment is performed. Two custom-made circuit boards equipped with DWM1000 UWB transceivers \cite{dw1000} are fixed to two static tripods as shown in Figure~\ref{fig:optimal_delay_experiment}. They are both connected to a Dell XPS13 computer running Ubuntu Desktop~20.04. 

First, a SS-TWR experiment is performed with 145 trials, for a total of 362500 measurements. \diff{Nothing varied in between trials, but the purpose of this experiment is to obtain the average rolling variance of SS-TWR experiments in order to get a value for $\sigma$, which is found to be $\sigma = 0.0682$ ns when averaging the variance over windows of 50 measurements. This value is used to plot the theoretical curves in Figure~\ref{fig:optimal_delay_metrics}. The need for computing a rolling variance rather than a single value for all measurements is because of the bias of SS-TWR measurements drifting over time due to the time-varying clock skew. Nonetheless, it is worth mentioning that knowing exactly the value of $\sigma$ is not necessary to perform the optimization in \eqref{eq:opt}, as finding the optimal delay requires solving \eqref{eq:optimal_delay}, which is independent of $\sigma$.}

With knowledge of the derived theoretical curves and the value of $\sigma$, the DS-TWR experiments are then performed to validate these values. The second-response delay $\Delta t^{53}$ is varied in between many trials, and for each trial 2500 measurements are collected to compute the average variance and rate for that specific value of $\Delta t^{53}$. The results for two different values of $\Delta t^{32}$ are shown in Figure~\ref{fig:optimal_delay_metrics}, where $\rho = 7.2$ ms is found experimentally to be the time required by the computer to process a range measurement. The experiment with $\Delta t^{32} = 0.35$ ms involves 5000 trials for a total of 12.5 million measurements, while the experiment with $\Delta t^{32} = 2$ ms involves 6600 trials for a total of 16.5 million measurements.

Given that this is a static experiment, $R_\mathrm{avg}$ essentially represents the variance in the measurement obtained by averaging out all recorded measurements over a span of one second. Crucially, both experiments presented here match the theoretical expectations quite well. As $\Delta t^{53}$ increases, both the standard deviation and the rate of the measurements decrease, and the optimal $\Delta t^{53}$ can then be found by finding the value that minimizes $R_\mathrm{avg}$. The experimental minimum does match the theoretical minimum, thus motivating the presented analytical optimization problem \eqref{eq:opt}. Lastly, as expected, the experiments with a longer $\Delta t^{32}$ have an order of magnitude higher standard deviation in the measurements, and in both experiments the standard deviation decreases as $\Delta t^{53}$ increases.

\section{Conclusion} 
This paper extends the comparison of SS-TWR and DS-TWR to include precision by deriving an analytical model of the variance \ms{and the CRLB} as a function of the signal timings of the ranging protocols. This consequently allows optimizing over the timing delays in order to minimize the variance of DS-TWR measurements, and an optimization problem is then formulated to maximize information by balancing the effect of reduced variance and reduced rate of measurements as timing delays increase. It is also shown that the effect of motion is independent of the timing delays in the utilized DS-TWR protocol. Lastly, the analytical variance model and optimization procedure are evaluated on an experimental set-up with two static ranging UWB transceivers. \ms{Future work will address finding optimal delays when the ranging protocol is customizable beyond standard DS-TWR, or when new TWR instances can be initiated before others are done.}

\correspauthor%

\bibliographystyle{IEEEbib}
\bibliography{ref}

\end{document}